\documentclass{emulateapj}\usepackage{psfig,graphicx,natbib}

\shorttitle{Estimating Stellar RV Variability from Kepler and GALEX}
\shortauthors{Cegla et al.}
\begin{document}
\title{Estimating Stellar Radial Velocity Variability from Kepler and GALEX:\\ Implications for the Radial Velocity Confirmation of Exoplanets}
\author{H.~M. Cegla\altaffilmark{1,2}, K.~G. Stassun\altaffilmark{2}, C.~A. Watson\altaffilmark{1}, F.~A. Bastien\altaffilmark{2}, J. Pepper\altaffilmark{2,3}}
\altaffiltext{1}{Astrophysics Research Centre, School of Mathematics \& Physics, Queen's University, University Road, Belfast BT7 1NN, UK; {\tt hcegla01@qub.ac.uk}}
\altaffiltext{2}{Department of Physics \& Astronomy, Vanderbilt University, Nashville, TN 37235, USA}
\altaffiltext{3}{Physics Department, Lehigh University, Bethlehem, PA 18015, USA}

\begin{abstract}
We cross-match the GALEX and Kepler surveys to create a unique dataset with both ultraviolet (UV) measurements and highly precise photometric variability measurements in the visible light spectrum. As stellar activity is driven by magnetic field modulations, we have used UV emission from the magnetically heated gas in the stellar atmosphere to serve as our proxy for the more well-known stellar activity indicator, ${R'}_{HK}$. The ${R'}_{HK}$ approximations were in turn used to estimate the level of astrophysical noise expected in radial velocity (RV) measurements and these were then searched for correlations with photometric variability. We find significant scatter in our attempts to estimate RV noise for magnetically active stars, which we attribute to variations in the phase and strength of the stellar magnetic cycle that drives the activity of these targets. However, for stars we deem to be magnetically quiet, we do find a clear correlation between photometric variability and estimated levels of RV noise (with variability up to $\sim$ 10 m~s$^{-1}$). We conclude, that for these quiet stars, we can use photometric measurements as a proxy to estimate the RV noise expected. As a result, the procedure outlined in this paper may help select targets best-suited for RV follow-up necessary for planet confirmation.
\end{abstract}


\section{Introduction}
\label{sec:intro}
The detection and confirmation of exoplanets today is dependent on our knowledge of the host star. This knowledge must go beyond solely the host star mass, and must also include an understanding of the star's inherent activity, i.e. astrophysical noise or `stellar jitter'. Astrophysical noise may arise from any phenomena inherent to the host star that alters the shape of the stellar absorption lines, injecting spurious or systematic RV signals, and/or altering the intensity of a transit light curve. Such phenomena include, but are not limited to, starspots, faculae/plage, granulation, stellar oscillations, and even potentially variable gravitational redshift. Ignoring or misunderstanding astrophysical noise for magnetically active stars can lead to stellar activity signals masking or mimicking planets; a problem arising for not only low-mass, low-amplitude planets, but also even higher mass planets such as hot Jupiters (e.g \citealt{queloz01}; \citealt{desidera04}; \citealt{huelamo08}; \citealt{huerta08}; \citealt{figueria10}). One method to avoid these false positives is to target more magnetically quiet stars.  

In this paper, we illustrate a relationship between photometric noise and predicted radial velocity (RV) noise. Such a relationship creates a filtering mechanism to prioritise planetary candidates in transit surveys that are ideal for RV follow-up observations (i.e. systems with magnetically quiet host stars), necessary for most planet confirmations. From targets in the GALEX survey, \cite{findeisen11} have demonstrated a potential empirical relationship between stellar ultraviolet (UV) emission and chromospheric activity (as indicated by ${R'}_{HK}$). In addition to this, there are also other empirical relationships between chromospheric activity and the predicted level of RV noise \citep{santos00,saar03,wright05, martizen-arnaiz10}. 

Unfortunately, targets in transiting exoplanet surveys do not always have UV and/or ${R'}_{HK}$ measurements available to predict the level of RV noise. However, they will have measurements of photometric variability, thus we use this parameter to predict RV noise. Due to its extreme precision and overlap with the GALEX survey, we used photometric variability measurements, in visible light, from Kepler to establish a relationship between photometric noise and UV emission. We find a significant correlation between photometric variability and Far-UV (FUV) emission, link it with chromospheric activity, and ultimately estimate the level of RV noise to determine if RV follow-up, and therefore planet confirmation, is feasible. As low-mass planets exhibit sub m s$^{-1}$ signals, it is prudent to know which stars are truly RV quiet.   

We discuss how the target sample in this paper was obtained in Section~\ref{sec:targets}, while the various data (photometric variability, UV emission, ${R'}_{HK}$ and RV noise estimates) for our sample is further discussed in Section~\ref{sec:kg_data}. UV excess and ${R'}_{HK}$ are analysed as stellar activity indicators in Section~\ref{sec:analysis_activity}, and used to estimate and analyse RV noise in Section~\ref{sec:RVanalysis}. Finally, we discuss our findings and conclusions in Section~\ref{sec:kepgal_disc}.

\section{The Target Sample}
\label{sec:targets}
Our target sample contains a variety of data from various sources. The Kepler Quarter 9 long cadence ($\sim$ 30 min exposures; \citealt{borucki10}) dataset forms the basis of our sample (this quarter was chosen for a direct comparison with \cite{bastien13}, see Section~\ref{subsec:phot}). The Kepler survey provides an ideal dataset for our variability studies due to its unprecedented photometric precision, large sample size, and overlap with the GALEX survey (see \citealt{martin05} for mission details). The overlap with the GALEX survey is of utmost importance to our study as it provides UV measurements for our targets. As a result, the $\sim$150,000 targets from the Kepler Input Catalog (KIC; \citealt{brown11}) were cross-matched with the GALEX survey (using the KGGoldStandard\footnote{http://archive.stsci.edu/kepler/catalogs.html }), leaving $\sim$ 30,000 targets with both Kepler and GALEX statistics. This was then further constrained to only those targets with GALEX FUV measurements (discussed in Section~\ref{subsec:fuvrhk}), limiting the sample to just under $\sim$ 4,000 targets (this reduction is most likely due to the faintness of the majority of Kepler targets, which makes the detection of FUV emission difficult). This dataset was also cross-matched with the Sloan Digital Sky Survey (see \citealt{york00} for SDSS details), which provided measurements of $u', g', r', i'$, that were later converted to V and B magnitudes following \cite{smith02}. Additional cross-matching with the 2MASS survey provided J and K magnitudes (one target was removed that had a 2MASS J - K colour $< $ -0.5). Later in the paper, this target sample is further restricted in keeping with \cite{bastien13} and \cite{findeisen11} as we use defining relationships from both of these authors, see Sections~\ref{subsec:phot} and \ref{subsec:fuvrhk} for details. 

\section{Data}
\label{sec:kg_data}
Throughout this section we outline the photometric variability measurements from Kepler and summarise the findings of  \cite{bastien13}. We also define our UV activity measurement, FUV excess, as well as our conversions from FUV to ${R'}_{HK}$, and later the conversion to estimated RV noise. 

\subsection{Photometric Variability Measurements}
\label{subsec:phot}
For Kepler targets, there are numerous photometric variability measurements available, for a large set see \cite{basri11}. Using three of these measurements, range (R$_{\rm var}$), number of zero crossing (X$_0$), and `8-hr flicker' ($F_8$), \cite{bastien13} were able to construct an evolutionary diagram of brightness variability, which we will use to explore the target sample in this paper. The range is a measurement of the large amplitude variation over the quarter (90 days), and is used as a tracer for spot modulation. The number of zero crossings is determined by smoothing the light curve by 10 hr bins and counting the number of times the resultant light curve crosses its median value; it provides an estimate of the complexity of the light curve and is more sensitive to higher frequency signals such as those from granulation. The 8-hr flicker is the root-mean-square (rms) on timescales shorter than 8 hours (rms of the residual light curve after a 16-point (8-hr) boxcar smoothing has been subtracted off the original). We have corrected the 8-hr flicker for dependencies on stellar brightness using the formulation given in the Supplementary Material of \cite{bastien13}. 

In \cite{bastien13}, the authors construct their evolutionary diagram of brightness variability by plotting R$_{\rm var}$ against $F_8$ with symbol size indicating X$_0$ and colour based on asteroseismic measurements of surface gravity (log g). This diagram consists largely of two populations: a vertical cloud (in R$_{\rm var}$) of dwarf stars with low values of both X$_0$ and $F_8$, as well as a horizontal sequence  of targets with low values of R$_{\rm var}$, large values of X$_0$, and increasing values of $F_8$ as surface gravity decreases. The authors speculate that the dwarf stars with low $F_8$ have a large spread in R$_{\rm var}$ from star to star due to the presence of targets in their sample that are both magnetically active, fast spinning, young, spotty dwarfs (those with large amplitude modulations and simple light curves, hence high R$_{\rm var}$ and low X$_0$ ) and those stars that are further in their evolution and have begun to spin down and decrease in stellar activity. There may also be dwarf stars that are intrinsically magnetically quiet (such as those in a Maunder minimum state) and/or spotty stars with varying levels of inclination. \cite{bastien13} argue that as these magnetically active, young dwarfs spin down they will eventually alight onto the horizontal sequence  they term the `flicker floor'. Once on the flicker floor, targets are believed to remain on the floor and continue to move towards higher values of $F_8$ as they evolve towards the red giant phase; this is because once stars have spun down the dominant astrophysical noise source is due to granulation (a complex signal, thus large X$_0$) and as stars evolve their granular cells increase in diameter and the convective turnover timescale increases, resulting in an increase in $F_8$ since the granulation noise is no longer averaged out over 8-hr timescales. 

In this paper, we have used these three photometric measurements, and knowledge of the stars' evolutionary state from the analysis of \cite{bastien13}, to explore their relationship with UV emission. In keeping with these authors, we only include targets that meet the following additional constraints: Kepler magnitudes of 7-14 (largely to avoid very faint targets), periods (G. Basri, private comm. -- see \cite{walkowicz13} for potentially improved rotation periods) of 3-45 days (to avoid very fast rotating stars and/or short-period pulsators and to avoid over extrapolation of period estimates as KIC data in our sample only spans 90 days), temperatures (as determined from KIC) of 4500-6650~K (to avoid very cool and hot stars), and X$_0$ per day between 0.01 and 2.1. With these additional constraints there are 944 stars remaining in the target sample.

\subsection{Activity Indicators: UV Excess and ${R'}_{HK}$}
\label{subsec:fuvrhk}
The overlap between the Kepler and GALEX surveys provides a means to estimate stellar activity from both visible light variations and UV emission. As stellar activity is driven by magnetic field interplay, measurements of the UV emission from magnetically heated gas located above the visible light photosphere can also serve as a proxy for magnetic activity and ultimately astrophysical noise. Additionally, as a star ages it begins to spin down, decreasing the interaction between the radiative and convective zones and consequently the stellar surface magnetic flux. This drop in magnetic flux leads to a decrease in both stellar activity and UV emission. Thus, measurements of UV emission may also serve as tracers of stellar evolution \citep{findeisen11}. Combining knowledge of the UV emission with visible light variations provides a more complete picture of the underlying physics. Further to this, there are also tentative relationships, from \cite{findeisen11}, between UV emission and the more well known optical stellar activity indicator, ${R'}_{HK}$.

\cite{findeisen11} have shown that stars with an excess of UV emission have higher stellar activity. In this paper we focus only on far-UV (FUV) emission for our stellar activity proxy as it was found to have a clearer relationship with ${R'}_{HK}$ (though this effect is present in both near-UV and far-UV). The FUV magnitudes for our final target sample (with the above constraints) range from 15.7 -- 24.4 and as a result are expected to be within the linear response region of the GALEX detector, so no signal-to-noise ratio cuts or further limiting magnitude cuts were made. \cite{findeisen11} have defined FUV excess to be the difference between the measured FUV flux and that predicted by the Kurucz photospheric models\footnote{http://kurucz.harvard.edu/}, normalised by the bolometric flux. For this study, we do not have bolometric magnitudes nor estimates of the underlying FUV present in the photosphere and as a result cannot measure a normalised FUV excess as in \cite{findeisen11}. As such, we follow the earlier work in \cite{findeisen10} and estimate FUV excess based off colour estimates in the FUV - J$_{\rm mag}$ vs J-K colour plane. The stellar locus of the colour diagram is then used to define the FUV excess as the difference between the measured UV and this locus. Following \cite{findeisen10}, we define our cut-off for FUV excess to be
\begin{equation}
\label{eqn:fuvX}
FUV - J = 23.5(J- K) + 4.35,
\end{equation}
where J is the 2MASS J magnitude and J-K is the 2MASS JK colour, and all data lying below this cut-off are deemed to have FUV excess (more negative values indicate larger excess). It is important to note that the main difference between our definition of FUV excess and that of \cite{findeisen10} is the position of the zero point for FUV excess (the source of which may be due to differences in extinction that have not been accounted for in this study). Furthermore, as FUV excess is only used in this paper to view the gradient of UV excess changes, and not to define an absolute scale (since we do not have normalised FUV excess values), this difference does not affect our studies. The reader may also see \cite{smith10} for an alternate definition of FUV excess using B-V colour instead of J-K. For this paper we prefer the J-K based definition of \cite{findeisen10} because of the availability of reliable J-K colours from 2MASS for nearly all stars in our GALEX sample, and because it does not rely on ${R'}_{HK}$ measurements.

In addition to identifying UV excess as an activity indicator, \cite{findeisen11} also used the UV measurements to obtain a preliminary relationship with the optical activity indicator ${R'}_{HK}$. ${R'}_{HK}$ (the fraction of a star's bolometric luminosity emitted in the Ca II lines due to chromospheric activity e.g. \citealt{noyes84}) is one of the most commonly measured, understood, and standardised stellar activity indicators. For these reasons, as well as a desire to eventually estimate RV noise, we have employed the empirical relationship between FUV and ${R'}_{HK}$ from \cite{findeisen11}, given below in Equation~\ref{eqn:Rhk},  

\begin{eqnarray}
\label{eqn:Rhk}
        \lefteqn{ {\rm log}~R'_{HK} =  -4.25 - 0.44(FUV - V - 12) } \nonumber \\
	&~~~~~~~~~+3.48(B - V - 0.8) \nonumber \\
	&~~~~~~~~~-0.35(B - V -0.8) \nonumber \\
	&~~~~~~~~~\times(FUV -V - 12),  \nonumber \\
	&{ \rm valid~over~the~ranges:} \nonumber \\
	&(0.5 \le B - V \le 0.7) \nonumber \\
	& {\rm and} \nonumber  \\
	&(3.4 \le FUV - V -12(B - V) \le 5.2),
\end{eqnarray}
where the V and B magnitudes are converted from the KIC catalog $u', g', r', i'$ magnitudes following \cite{smith02}, and the validity range provided represents the parameter space over which \cite{findeisen11} apply this relation (to control colour dependence and saturation effects). We note that this conversion to ${R'}_{HK}$ is based on an empirical relationship, unnormalised by the stellar photosphere and independent of any expected value of FUV flux inherent in the stars. As UV excess is a stronger tracer of stellar activity than UV emission alone, it would be preferred to use normalised FUV excess values to convert directly to ${R'}_{HK}$, however, this is not possible in this study for the reasons discussed previously. Further to this, we stress that this relationship be used with the utmost caution as \cite{findeisen11} find large scatter in their correlation between UV and ${R'}_{HK}$, and as a result it cannot be used to determine precise measurements. However, use of this relationship is acceptable to examine the overall nature of stellar activity within our sample and is used in Section~\ref{subsec:RVnoise} to obtain a very general estimate of the behaviour of RV noise. 

\begin{table}[b!]
\caption[Estimating RV Noise from ${R'}_{HK}$]{Coefficients for conversion from ${R'}_{HK}$ to RV Noise: $ \sigma_{RV}'\propto a10^5{R'}_{HK}^b$}
\centering
\begin{tabular}{c||ccc||ccc}
    \hline
    \hline
      &  & $a$ &  &  & $b$ &  \\
     Temperature (K) & Wright & Saar &  Santos & Wright & Saar &  Santos  \\
    \hline  
    $\ge$ 6000 \& $< 6650$ & 1 & 1 &  9.2 & 1.6 & 1.7 &  0.75\\
    $\ge$ 5200 \& $< 6000$ & 1 & 1 &  7.9 & 0.8 & 1.1 &  0.55\\
    $<$ 5200 & 1 & 1 &  7.8 & 0.8 & 1.1 &  0.13\\
    \hline
  \end{tabular}
\label{tab:rhk}
\end{table}

\subsection{Estimating Radial Velocity Noise}
\label{subsec:RVnoise}
The ultimate goal of this work is to use photometric variability measurements to estimate the amount of astrophysical noise expected in RV follow-up. As there are several empirical relations between ${R'}_{HK}$ and predicted RV noise (jitter) in the literature, we have chosen to explore those from \cite{wright05}, \cite{saar03} and \cite{santos00} as their relations are directly comparable to one another. Although \cite{wright05} employ a slightly different fitting technique and metric, they do offer a comparison to \cite{saar03}, which we have used in this paper for the same purpose. In all three papers, the authors find RV noise, hereafter denoted by $\sigma_{RV}'$, proportional to $a{R'}_{HK}^b$, values of $a$ and $b$ can be found in Table~\ref{tab:rhk} for the three different temperature regions used in this paper: 6650 - 6000~K, 6000 - 5200~K, $<$ 5200~K. Each of these relations provides similar results, differing in the absolute magnitude of predicted RVs (relationships from \cite{wright05} and \cite{saar03} predict  $\sigma_{RV}'$ $\approx$ 0-10 {\rm m~s$^{-1}$}, while \cite{santos00} predicts $\sim$ 5-25 {\rm m~s$^{-1}$}), but the overall behaviour and trends of the RVs remain the same. As we are interested in the general behaviour of  $\sigma_{RV}'$ rather than exact values, for ease of display, we have chosen to present  $\sigma_{RV}'$ estimates only from \cite{saar03}. In the next sections, we estimate ${R'}_{HK}$ from FUV, and convert ${R'}_{HK}$ to  $\sigma_{RV}'$. Finally, we relate  $\sigma_{RV}'$ to photometric variability. 

\section{Analysis of Activity Indicators}
\label{sec:analysis_activity}
\begin{figure}[b!]
\centering
\includegraphics[width=8.5cm]{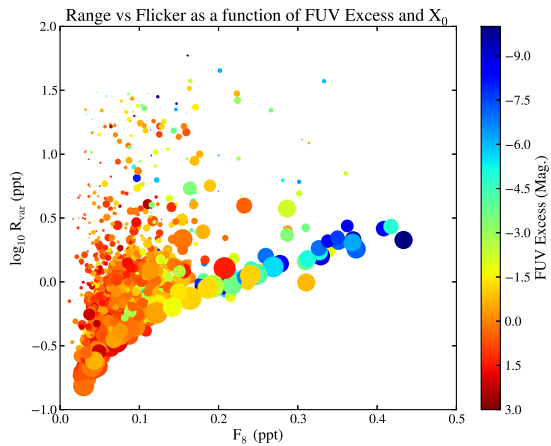}
\caption[Photometric Variability (R$_{\rm var}$ and $F_8$) as Function of FUV Excess]{Large amplitude modulation measurement (R$_{\rm var}$) vs rms $< $8 hr ($F_8$) as a function of the number of zero crossings (X$_0$) per day, i.e. complexity of light curve (symbol size), and amount of FUV excess (colour). Targets with the highest R$_{\rm var}$ also tend to have higher values of FUV excess. Additionally, within the flicker floor, we see a gradient in FUV excess where larger flicker tends to larger FUV excess. Note that stars with $F_8 > $ 0.5 do exist in the target population, but are not shown here as they skew the display.} 
\label{fig:fuvX}
\end{figure}

The estimates of FUV excess, from Equation~\ref{eqn:fuvX}, for our cross-matched sample of Kepler and GALEX targets were used to explore the `flicker sequence' of stellar evolution seen in \cite{bastien13}. In Figure~\ref{fig:fuvX}, we show the relationship of R$_{\rm var}$ against flicker ($F_8$) as a function of X$_0$, magnitude represented by symbol size, and amount of FUV excess, represented by colour. As expected, we find that targets with the highest values of R$_{\rm var}$, likely young spotty stars, have a higher FUV excess compared to those with low R$_{\rm var}$. However, we also found a FUV excess gradient in the flicker floor sequence, where as stars evolve onto the red giant branch they experience a large increase in FUV excess. 

The origin of this FUV excess gradient is unclear; it is likely that these FUV excess measurements are over-estimated for evolved stars in our method due to the fact that that giant star chromospheres are optically thick compared to dwarf stars, resulting in enhanced FUV colours (mistakenly measured as an FUV excess in our method). At the same time, there are reports of chromospherically active, tidally-locked, short-period subgiants and Li-rich K giants (that may have engulfed a close brown dwarf or hot Jupiter companion). One such report comes from \cite{isaacson10}, who found approximately 10\% of their sample of subgiants to be strongly chromospherically active spectroscopic binaries. \cite{wright05} also found that subgiant targets may exhibit higher astrophysical noise than targets on the main-sequence, and speculate on the possibility that there may exist red giants that exhibit significantly larger amounts of stellar jitter than their main-sequence counterparts. In this paper, we regard the FUV excess estimates for the giant stars with caution, and we do not attempt to ascribe a true physical origin for these apparent excesses. Rather, we simply take advantage of the fact that, on the flicker floor, the apparent FUV excess does clearly correlate with the stellar flicker and therefore with the stellar surface gravity. This then allows us to use the apparent FUV excess as a reliable predictor of RV noise, even if the FUV proxy is not a true excess in the chromospheric sense for these stars.

\begin{figure}[b!]
\centering
\includegraphics[width=8.5cm]{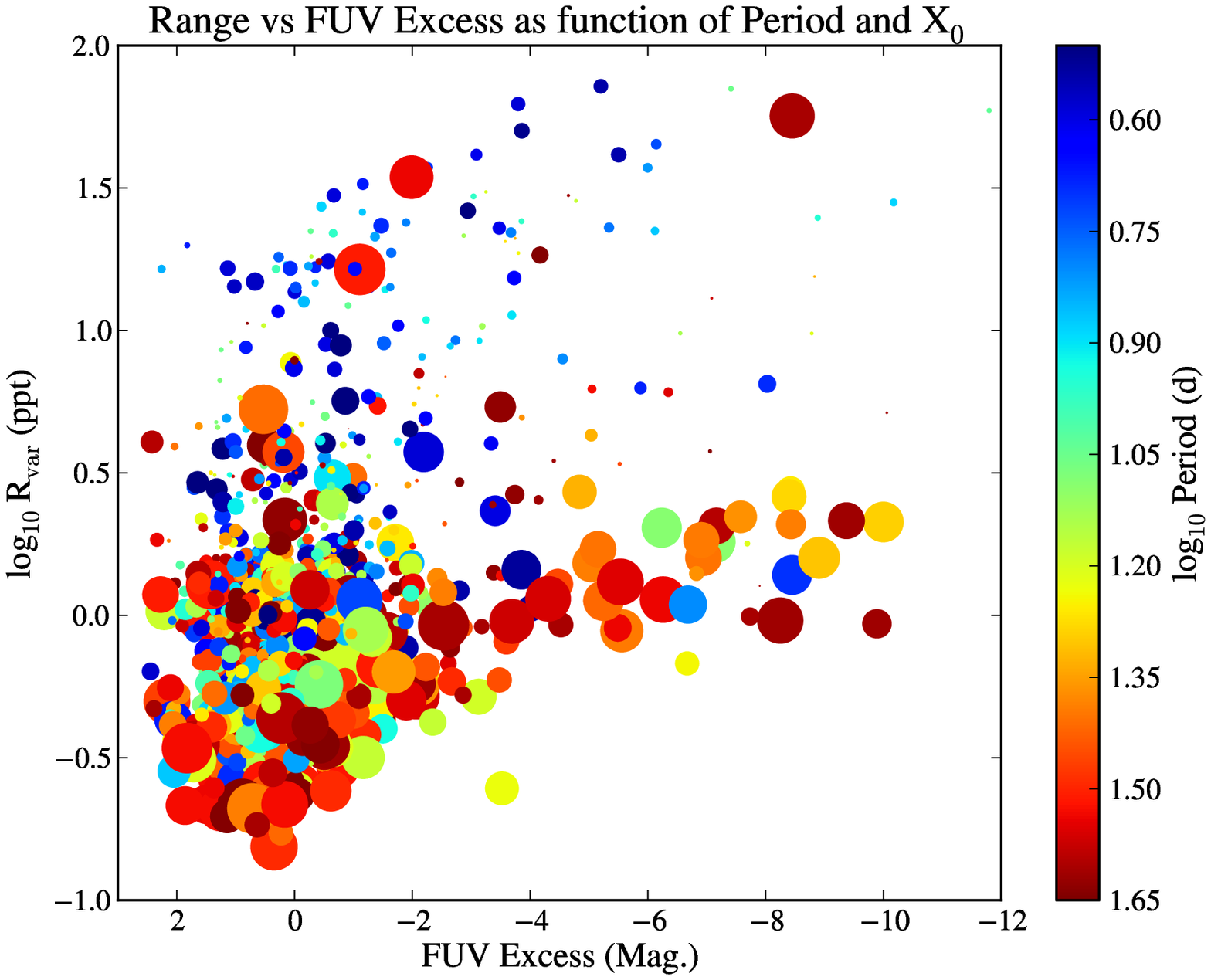}
\includegraphics[width=8.5cm]{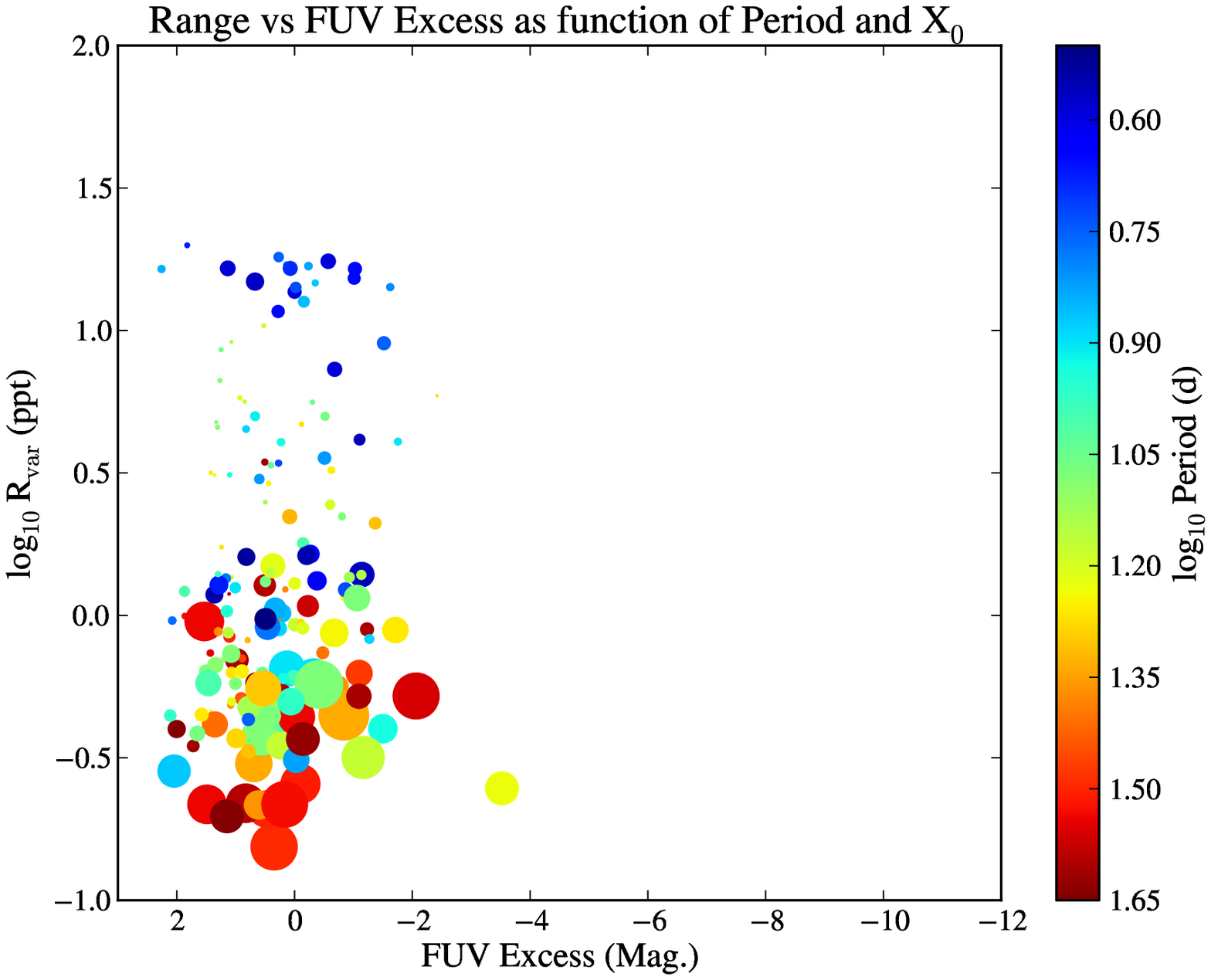}
\caption[Photometric Variability (R$_{\rm var}$) as a Function of FUV and Stellar Rotation Period]{Top: large amplitude modulation measurement (R$_{\rm var}$) vs FUV excess as a function of the number of zero crossings (X$_0$), i.e. complexity of light curve (symbol size), and stellar rotation period (colour). There is a distinct difference in rotation period, with fast rotators having much larger R$_{\rm var}$. Dwarf stars (largely the upper branch) with FUV excess also tend to be preferentially fast rotators with high amplitude modulation. There are also a couple of faster rotating giant stars (primarily the lower branch), which may be binaries. Bottom: same as above, but with the parameter space over which conversion to ${R'}_{HK}$ holds, given in Equation~\ref{eqn:Rhk} (targets with large $F_8$ were also removed at this stage).} 
\label{fig:fuvXper} 
\end{figure}

We also explored variations in stellar rotational period as a function of R$_{\rm var}$, FUV excess, and X$_0$ in the top of Figure~\ref{fig:fuvXper} (where R$_{\rm var}$ is plotted against FUV excess as a function of X$_0$ (symbol size) and period (colour)). We find a distinct difference between fast rotators, which have the highest amplitude variation (as expected for young, magnetically active stars) and slow rotators, which have lower amplitude variation. While a tentative relationship may be seen for the dwarf population (which largely compose the upper region and also some of the lower left of the top plot in Figure~\ref{fig:fuvXper}), where those targets with more FUV excess tend to have high R$_{\rm var}$ and fast periods (upper right of diagram), there are, however, targets with little or no FUV excess that still have high R$_{\rm var}$ and fast periods. There are also a couple of outliers to this, seen as slow rotators with complex light curves (large symbol size) in the upper region of this diagram; however, these targets have extremely low surface gravities (using the relationship between log g and  $F_8$ from \citealt{bastien13}), and should therefore likely be discarded. The lower right of this diagram, which we believe to be composed of giant stars, are all slower rotators, with the exception of a couple that may be binaries, with high values of FUV excess. The bottom plot in Figure~\ref{fig:fuvXper} displays the same data, but only for the parameter space over which the conversion to ${R'}_{HK}$ holds, and serves to display how significantly our sample size is affected by these limitations. However, keeping to this limited parameter space does remove the very evolved red giants where our FUV excess measurement may be systematically biased.  

\begin{figure}[b!]
\centering
\includegraphics[width=8.5cm]{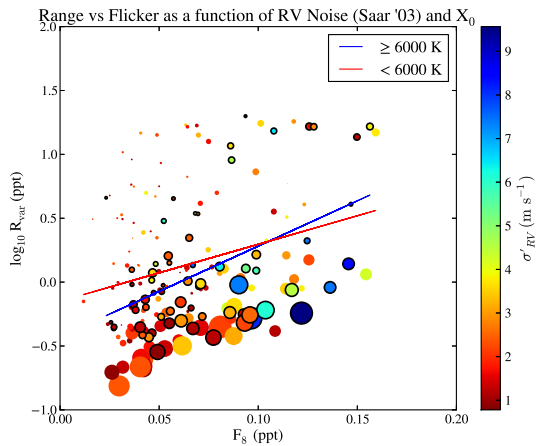}
\caption[Photometric Variability (R$_{\rm var}$ and $F_8$) as a Function of Estimated RV Noise]{Large amplitude modulation measurement (R$_{\rm var}$) vs rms $<$ 8 hr ($F_8$) as a function of the number of zero crossings (X$_0$) per day (symbol size), and level of predicted RV noise/jitter ($\sigma_{RV}'$; colour). Black perimeter indicates targets with temperature $>$ 6000~K. Linear fits between R$_{\rm var}$ and $F_8$ have been made to both hot and cool stars, shown in blue and red, respectively.} 
\label{fig:rangeflick_RV}
\end{figure}

\section{Results}
\label{sec:RVanalysis}
\subsection{Relationship Between Photometric Variability and RV Noise}
\label{subsec:phot_rv}
\begin{figure}[b!]
\begin{center}
\includegraphics[width=8.5cm]{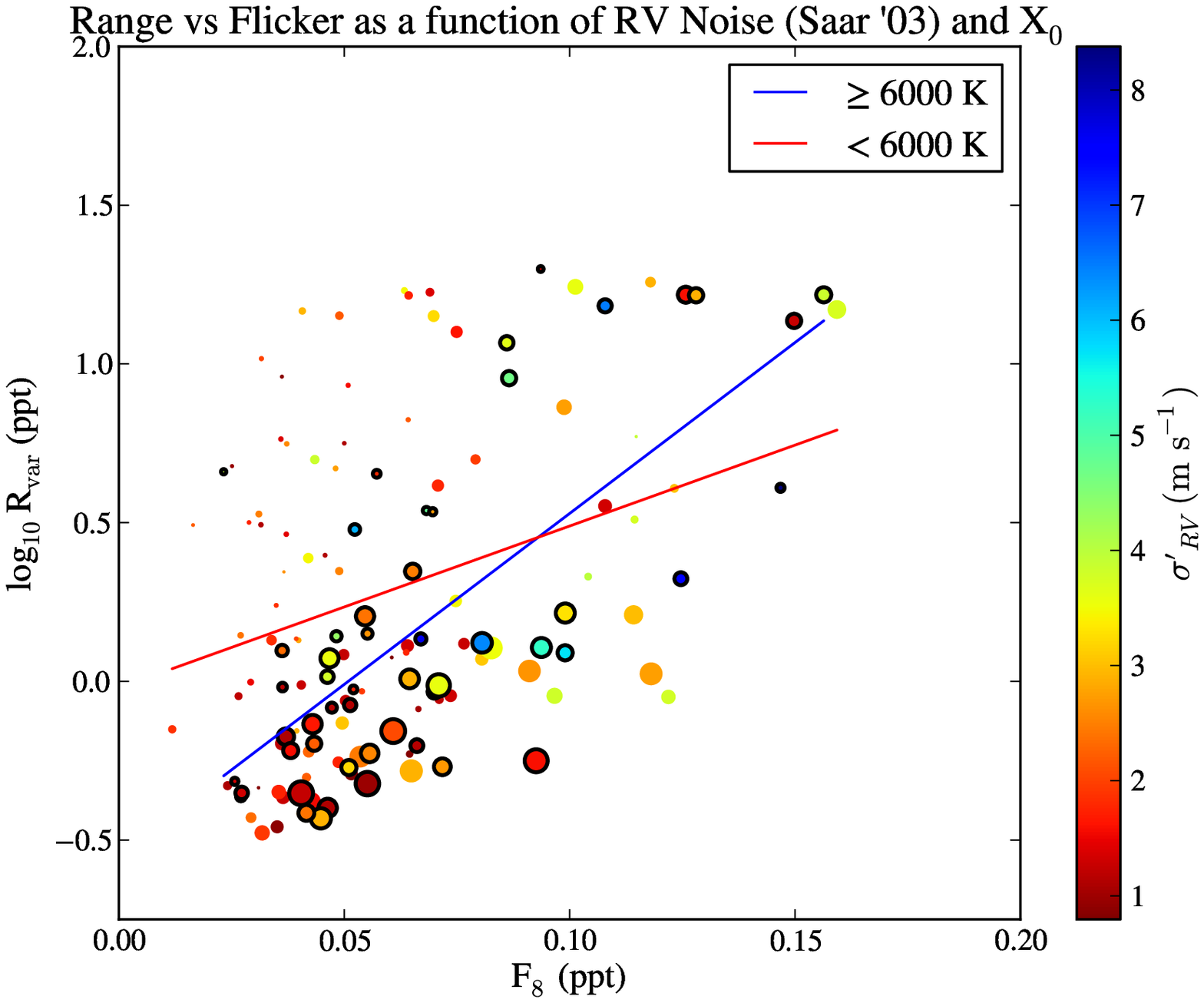}
\includegraphics[width=8.5cm]{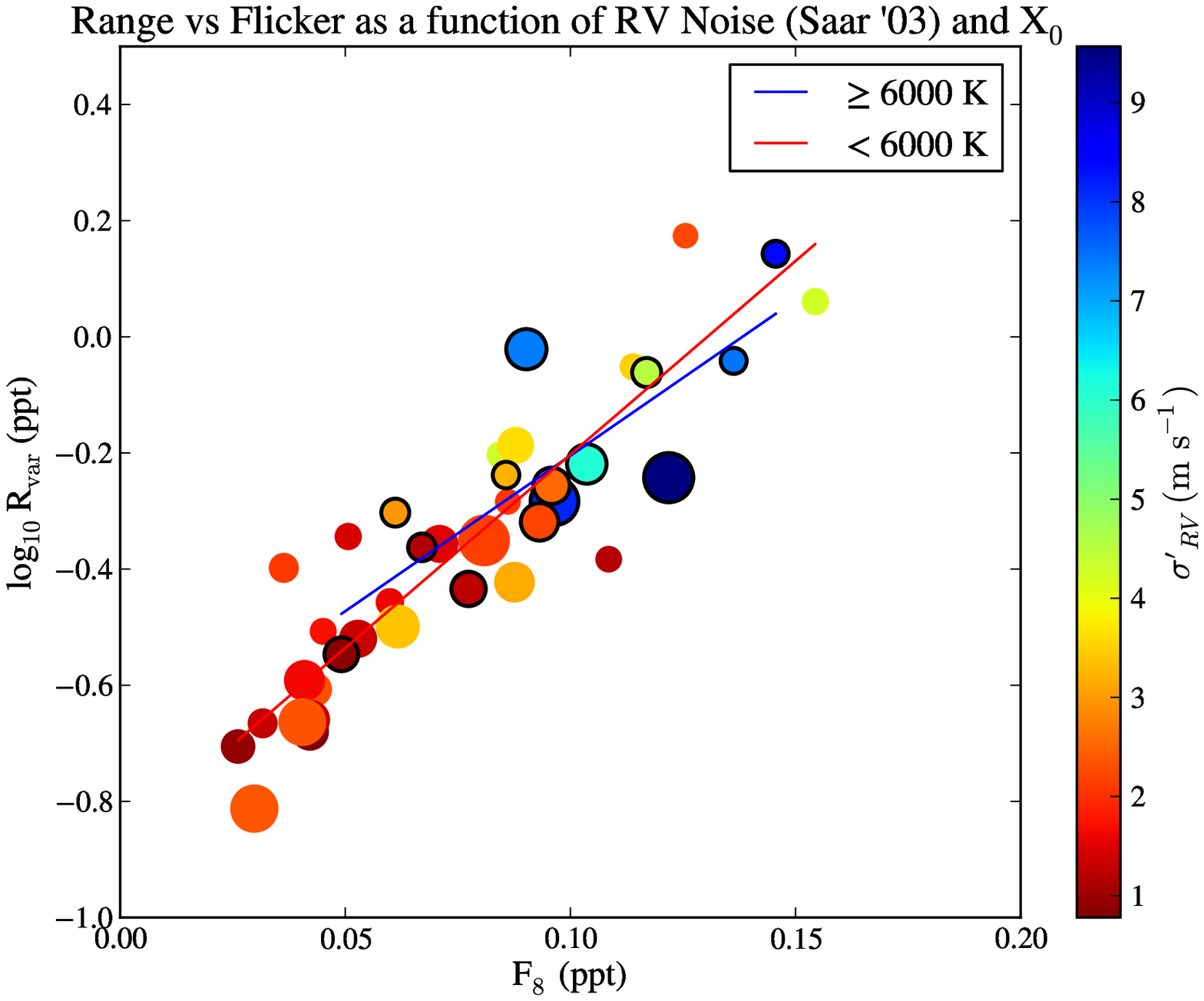}
\caption[Photometric Variability (R$_{\rm var}$ and $F_8$) as a Function of Estimated RV Noise for Active and Quite Stars]{Same as Figure~\ref{fig:rangeflick_RV}, but subdivided into those above the flicker floor (top) and those on the flicker floor (bottom). Those on the flicker floor are defined to be those stars that average more than one X$_0$ per day.}
\label{fig:flickerrange}
\end{center}
\end{figure}

The values of FUV emission for our target sample have been converted into an estimate of RV noise, $\sigma_{RV}'$, using the procedures described in Sections ~\ref{subsec:fuvrhk} and ~\ref{subsec:RVnoise}. In Figure~\ref{fig:rangeflick_RV}, again, R$_{\rm var}$ is plotted against $F_8$ as a function of X$_0$ (symbol size), but this time with colour now representing our estimate of  $\sigma_{RV}'$ and stars hotter than 6000~K depicted with a black perimeter. We find a $\sigma_{RV}'$ gradient is present in the flicker floor. However, for stars above the flicker floor, with higher R$_{\rm var}$, any gradient in $\sigma_{RV}'$ is less apparent, if present at all. We have performed a linear least squares fit between R$_{\rm var}$ and $F_8$, as an initial estimate, for stars both hotter and cooler than 6000~K separately as predicted RV noise varies depending on the temperature of the host star (often depicted by spectral type). Beyond 6000~K the mass of the convection zone decreases drastically, which ultimately affects the nature of the stellar activity. The linear fits between R$_{\rm var}$ and $F_8$ are quite poor and as a result, in Figure~\ref{fig:flickerrange}, we have further subdivided the data into flicker floor and non-flicker floor populations with the hypothesis that magnetically quiet stars (believed to be those on the flicker floor) may behave differently from more active stars; here the flicker floor is defined to be those stars that average more than one X$_0$ per day. 

\begin{figure}[b!]
\begin{center}
\includegraphics[width=8.5cm]{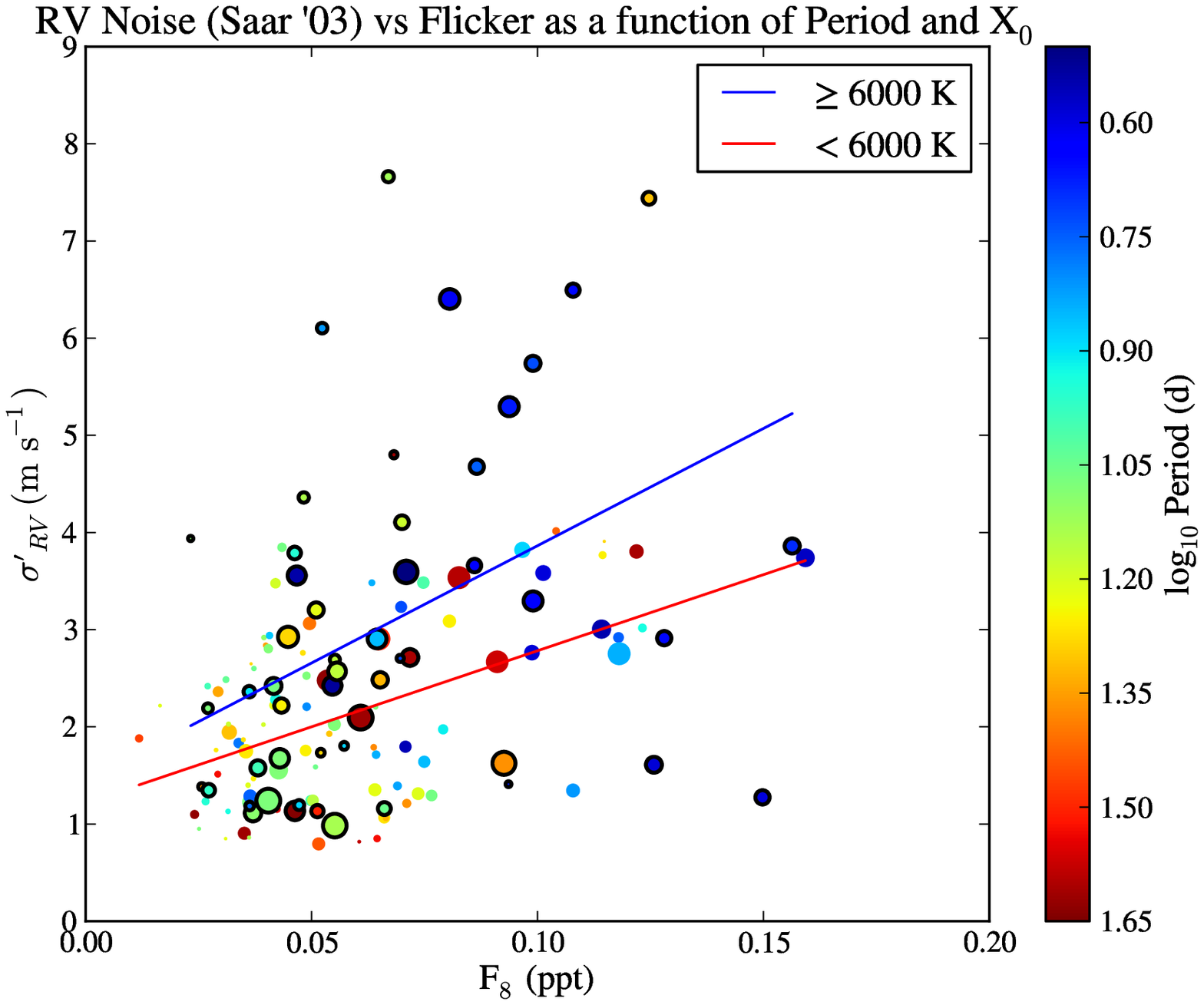}
\includegraphics[width=8.5cm]{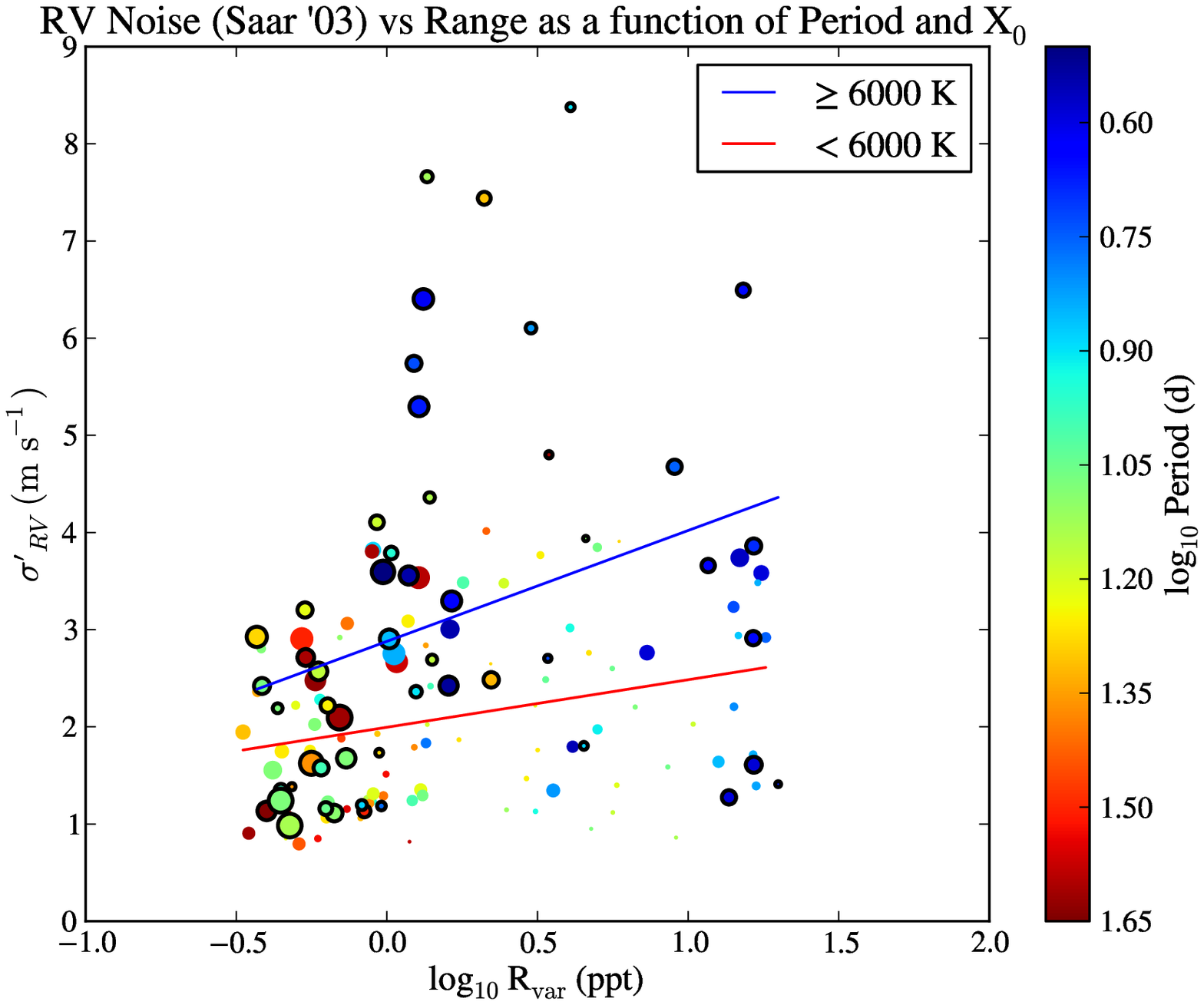}
\caption[Photometric Variability (R$_{\rm var}$ and $F_8$) as a Function of Estimated RV Noise for Active Stars]{Targets above the flicker floor: predicted RV noise ($\sigma_{RV}'$) vs photometric variability measurements $F_8$ (top) and R$_{\rm var}$ (bottom) as a function of X$_0$ (symbol size) and stellar rotation period (colour). Black perimeter indicates targets with temperature $>$ 6000~K. Linear fits have been made to both hot and cool stars, shown in blue and red, respectively.}
\label{fig:notfloor}
\end{center}
\end{figure}

For those stars above the flicker floor (top, Figure~\ref{fig:flickerrange}), there still lies significant scatter between both stars hotter and cooler than 6000~K, evidence that these stars can occupy a large parameter space in R$_{\rm var}$ and $F_8$. Though, if the stars are quiet both in range and flicker simultaneously then they are also quiet in $\sigma_{RV}'$, which is to be expected. The hotter stars do have a steeper gradient than the cooler stars, which may be due to the fact that hotter stars have larger, brighter, faster, longer-lived granules with greater contrast to the intergranular lanes \citep{beeck13}, all of which contribute to larger photometric and spectroscopic noise in both magnetically active and quiet stars alike. At the same time, there are a larger portion of cooler stars that have high R$_{\rm var}$, which we would expect since these targets are often more magnetically active and more likely to have a large spot presence. On the other hand, stars on the flicker floor (bottom, Figure~\ref{fig:flickerrange}) show a much tighter correlation between R$_{\rm var}$ and $F_8$, and the gradient in $\sigma_{RV}'$ is clear, with little difference between the gradient of stars hotter and cooler than 6000~K. If \cite{bastien13} are correct that flicker traces log g, it is likely that what is seen in the flicker floor is simply that magnetically quiet stars are dominated by granulation noise and, as they continue to evolve, granulation noise increases due to increases in granule diameters and convective turnover times. Indeed, \cite{cranmer13} demonstrate that granulation is in fact the dominant driver of $F_8$ for stars on the flicker floor sequence.

\begin{figure}[b!]
\begin{center}
\includegraphics[width=8.5cm]{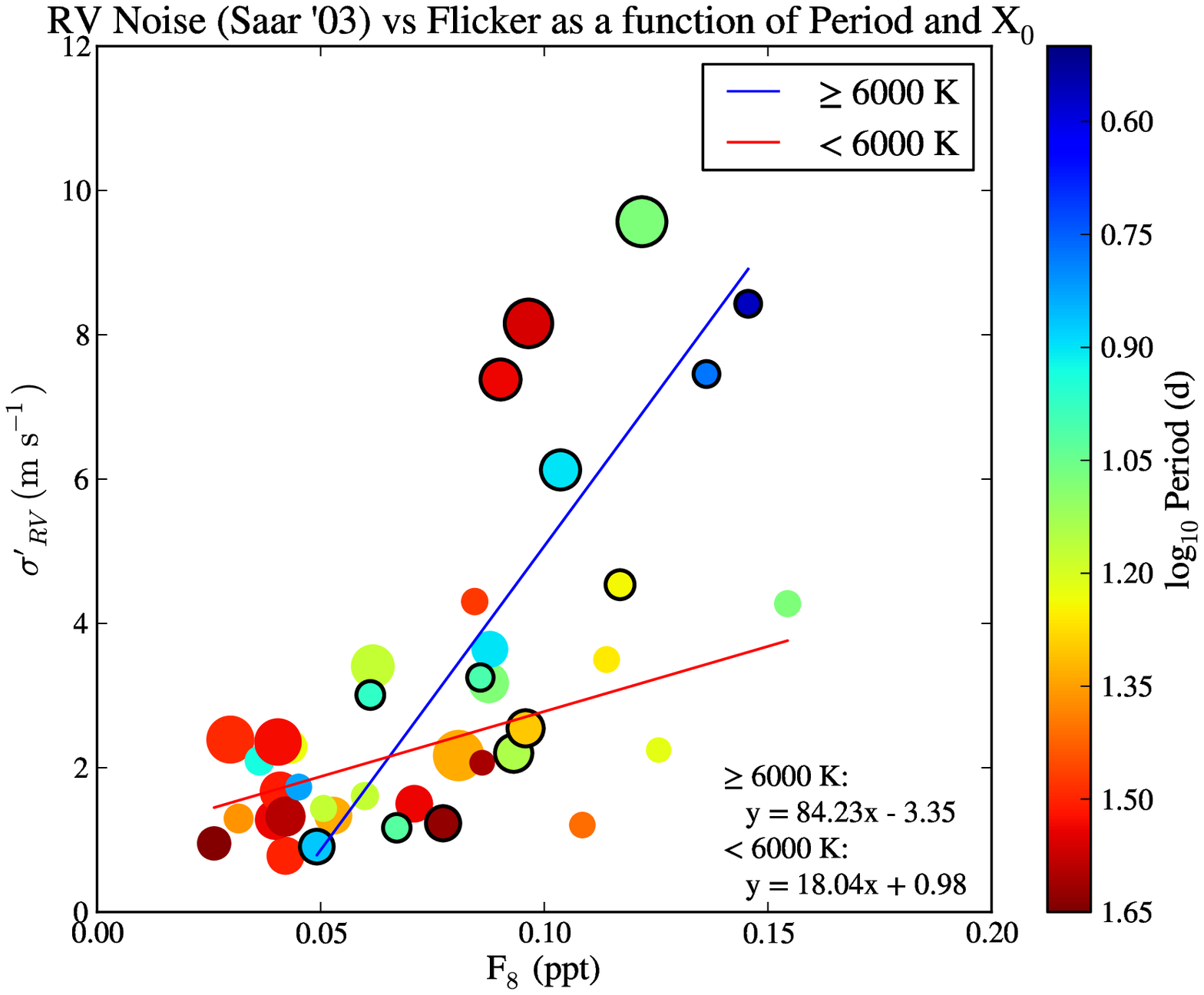}
\includegraphics[width=8.5cm]{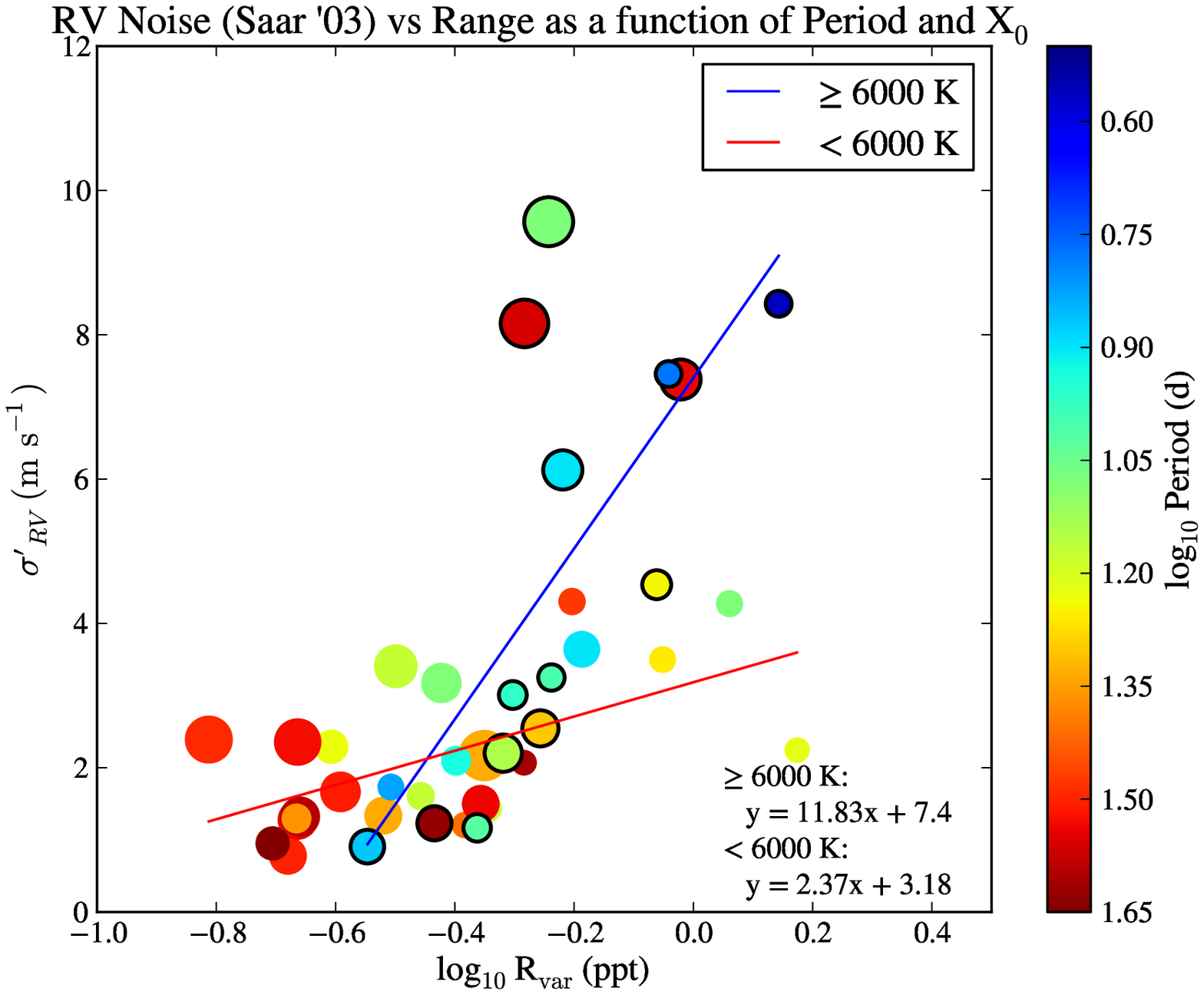}
\caption[Photometric Variability (R$_{\rm var}$ and $F_8$) as a Function of Estimated RV Noise for Quite Stars]{Targets on the flicker floor: predicted RV noise ($\sigma_{RV}'$) vs photometric variability measurements $F_8$ (top) and R$_{\rm var}$ (bottom) as a function of X$_0$ (symbol size) and stellar rotation period (colour). Black perimeter indicates targets with temperature $>$ 6000~K. Linear fits have been made to both hot and cool stars, shown in blue and red, respectively.}
\label{fig:floor}
\end{center}
\end{figure}

To further investigate correlations between $\sigma_{RV}'$ and these photometric measurements, we explored how the estimated values of $\sigma_{RV}'$ change as a function of R$_{\rm var}$ and $F_8$ separately, for both targets above and on the flicker floor, shown in Figures~\ref{fig:notfloor} and \ref{fig:floor}. For stars above the flicker floor (Figure~\ref{fig:notfloor}), even when examining R$_{\rm var}$ and $F_8$ separately, there is still significant scatter. For these stars, the dependence on period is more obvious, with faster rotators typically having larger $\sigma_{RV}'$, $F_8$ and R$_{\rm var}$ for both stars hotter and cooler than 6000~K. Though, there are instances of both high $F_8$ and R$_{\rm var}$ and lower levels of $\sigma_{RV}'$. This could be due to systematic differences in FUV between dwarfs and subgiants (subgiants according to \citealt{bastien13} have flicker between approximately 0.06 and 0.15), or could indicate that some of the RV signature of the spots and plage for these more active stars cancels out while the photometric signature does not. Though, if the latter were the case we would expect some inhibition of convection due to the presence of spots, the lack of apparent evidence for this in the flicker measurement could be due to longer convective turnover timescales for these targets masking any decrease in flicker due to higher magnetic activity. 

Stars on the flicker floor, again, show a much clearer picture of stellar activity. In Figure~\ref{fig:floor}, there is a clear correlation between estimated $\sigma_{RV}'$ and both R$_{\rm var}$ and $F_8$, which seems to be independent of rotation period. From our empirical fits in Figure~\ref{fig:floor}, we find

\begin{equation}
\label{eq:jitterF8f}
\sigma_{RV}'(\mathrm{m~s}^{-1}) = 84.23 \times F_8(\mathrm{ppt}) - 3.35 ~\mathrm{(\ge 6000~K)}
\end{equation}

\begin{equation}
\label{eq:jitterF8g}
\sigma_{RV}'(\mathrm{m~s}^{-1}) = 18.04 \times F_8(\mathrm{ppt}) + 0.98 ~\mathrm{(<  6000~K)}
\end{equation}

\begin{equation}
\label{eq:jitterrange}
\sigma_{RV}'(\mathrm{m~s}^{-1}) = 11.83 \times \mathrm{log_{10}R_{var}}(\mathrm{ppt}) + 7.4~\mathrm{(\ge 6000~K)}
\end{equation}

\begin{equation}
\label{eq:jitterrange}
\sigma_{RV}'(\mathrm{m~s}^{-1}) = 2.37 \times \mathrm{log_{10}R_{var}}(\mathrm{ppt}) + 3.18~\mathrm{(<  6000~K)}.
\end{equation}
The slopes and intercepts of these correlations are sensitive to both the $B-V$ colour and the temperature of the targets. This is because the conversion from FUV to ${R'}_{HK}$ in Equation~\ref{eqn:Rhk} uses $B$ and $V$ magnitudes, as well as the fact that the hot and cool stars are separated by a temperature cut-off. We have repeated the above analysis with the $B$ and $V$ magnitudes provided from \cite{everett12} and the updated temperatures provided from \cite{pinsonneault12}. The \cite{pinsonneault12} temperatures are systematically hotter than the KIC temperatures, and have been shown to also be systematically hotter than spectroscopically determined temperatures for stars hotter than 6000~K. Consequently, the cut-off from hot to cool stars in our target sample had to change from 6000~K to 6250~K. The results of these new updates was an increase in the Pearson's R correlation coefficient in Figure~\ref{fig:floor} for cool stars (0.6 to 0.67) and a decrease for hot stars (0.77 to 0.59), the slopes and intercepts changed slightly, but the overall picture remained the same. Since the \cite{pinsonneault12} temperatures are known to be systematically off for hotter stars, and we are attempting to differentiate between cooler and hotter stars, we opted not to use the \cite{pinsonneault12} temperatures in our final analysis. Additionally, since we have opted to use the KIC temperatures, we also have continued to use the $B$ and $V$ magnitudes converted from the $u', g', r', i'$ magnitudes following \cite{smith02}. The reason for this is because the temperatures in the KIC catalogue were calculated closely following \cite{smith02}, and we want the $B-V$ colour to be consistent with the temperature calculations. More accurately determined temperatures in the future are needed to clear up such discrepancies. 

As we would expect, the gradient in Equations~\ref{eq:jitterF8f} -- \ref{eq:jitterrange} is much steeper for the hotter stars which we attribute to the larger, brighter, faster, longer-lived granules with greater contrast to the intergranular lanes, resulting in more `vigorous' granulation as compared to the cooler stars. As these stars are believed to be magnetically quiet, the dominant noise source is likely granulation, and as such, Figure~\ref{fig:floor} may illustrate the dependence of granulation noise on its evolutionary phase and spectral type. The amount of RV noise estimated from these relationships is between $\sim$ 0.8 -- 10 m s$^{-1}$, which is in agreement with measured values for magnetically quiet stars \citep[e.g.][]{santos00,saar03, wright05}. 

The elevated levels of estimated RV noise as one moves along the flicker floor are presumably the result of increased levels of granulation. However, one must caution that this stems from a relationship between $\sigma_{RV}'$ and ${R'}_{HK}$, the latter being estimated from FUV-V. The reason for increased FUV flux relative to the V-band along the flicker floor (which drives the increasing trend in $\sigma_{RV}'$) is unclear. It may be that granulation does indeed result in increased FUV-V, yet one would expect that these stars are chromospherically quiet and as such have low levels of ${R'}_{HK}$. Therefore the relationship between FUV, ${R'}_{HK}$, and $\sigma_{RV}'$ may not be directly applicable in a regime where we think there is strong granulation but not strong chromospheric activity (i.e. evolved stars). As mentioned previously, however, chromospherically-active evolved stars have been reported and it may be that our relationships only apply for those stars. Nonetheless, it is clear that the FUV-V index for stars on the flicker floor traces stellar evolution (i.e., FUV-V on the flicker floor clearly correlates with log g), and therefore serves as a proxy for granulation noise, which appears in turn to correlate with RV noise. A key test of the relationships would be to obtain ${R'}_{HK}$ and RV variations of evolved stars.

\subsection{Testing the RV Noise Relationships Against the Sun}
\label{subsec:sub}
Although more detailed follow-up is necessary, we have tested the ability of the fits provided here, for quiet stars, to predict the RV variability of the quiet Sun. From the SOHO data presented in \cite{bastien13}, and references therein, it is clear that the Sun at its minimum (in its magnetic field cycle) is in close approximation to the flicker floor sequence. As such, we expect to be able to estimate the $\sigma_{RV}'$ at solar minima from only R$_{\rm var}$ and $F_8$. We find that using R$_{\rm var}$ =  0.1 ppt and $F_8$ = 0.015 ppt (the corresponding values for the solar minima), the predicted $\sigma_{RV}'$ at solar minima is between $0.8 - 1.3 \rm{~m~s}^{-1}$. These values are in agreement with the disc-integrated RV variability of the Sun as measured at the Mount Wilson Observatory for solar cycles 20 -- 21 \citep{makarov10}. 

On the other hand, these estimated values of $\sigma_{RV}'$ are higher than recent theoretical models. For example, \cite{dumusque11c} and \cite{meunier10a} report the expected solar RV variability to be $\sim$ $0.4 - 0.5 \rm{~m~s}^{-1}$. One such reason for this discrepancy could be that the Sun at solar minimum does not lie exactly on the flicker sequence and thus may not be strictly tied to the same behaviour as those stars that lie on the flicker floor. Additionally, the above relationships could be skewed for the quietest stars in this study due to the instrumental precision available for the empirical relationships between ${R'}_{HK}$ and $\sigma_{RV}'$. For example, the `quietest' stars may exhibit RV variability that lies below the RV precision achievable by the instrumentation. This would result in those stars (with low ${R'}_{HK}$) flatlining in their relationship with $\sigma_{RV}'$, leading to a systematic over-estimation of their true RV variability. More high precision RV noise measurements on stars similar to the Sun are needed before either of these can be known for certain. Nonetheless, these results are still in agreement to both observed and theoretical values to $<$ 0.5 $\rm{~m~s}^{-1}$, which should be sufficient to distinguish which targets are ideal for RV follow-up.

\section{Summary and Concluding Remarks}
\label{sec:kepgal_disc}
As space-based transit surveys with unprecedented photometric precision, such as Kepler and future missions (e.g. TESS or PLATO 2.0), unveil a plethora of potential exoplanet candidates, it is imperative that we understand the nature of the stellar hosts. The majority of these candidates require RV follow-up for planet confirmation, and due to the vast number of potential candidates determining which of these are best suited for follow-up will greatly increase planet confirmation efficiency. Further to this, a greater understanding of the stellar activity of the host star will also help prevent false positives in both magnetically active and quiet stars, and is most certainly necessary to confirm long-period, terrestrial planets around low-mass stars. 

Throughout this paper we have used FUV as a proxy for ${R'}_{HK}$ \citep{findeisen11}, which allowed us to estimate the amount of RV noise, $\sigma_{RV}'$, present for these stars \citep{saar03}. We have illustrated a distinct difference between stars on and above the flicker floor coined by \cite{bastien13}. Those on the flicker floor are targets lacking large amplitude modulations (hence, low R$_{\rm var}$), have more complex light curves (high X$_0$), and are believed to be magnetically quiet with granulation the dominant noise source. Those above the flicker floor, on the other hand, are more magnetically active and occupy substantial parameter space in the large amplitude modulations, likely due to varying spot presence. 

Targets that lie above the flicker floor show a breadth of variability in both photometric measurements and predicted $\sigma_{RV}'$. There is some evidence for a general trend that the faster rotating targets with higher levels of large amplitude modulations (spotty stars with high R$_{\rm var}$) in this sample have larger $\sigma_{RV}'$, but there is significant scatter in this and several targets have high R$_{\rm var}$ and flicker, with low $\sigma_{RV}'$ estimates. It is possible that some of the scatter seen in the data may be due to inclination effects and/or the behaviour of subgiants as compared to their main-sequence counterparts (including inherent FUV in the stellar photosphere, which could effect our $\sigma_{RV}'$ estimates). These stars could also very well be at different phases in their magnetic activity cycle, which would account for varying levels of spot activity and hence contribute to the scatter seen in R$_{\rm var}$. For more magnetically active stars, more data (i.e. information on stellar inclination and contaminating binaries, as well as actual RV and ${R'}_{HK}$ measurements, and more accurate colours and temperatures) over a larger parameter space is needed before we can conclude whether we can estimate $\sigma_{RV}'$ from photometric measures of R$_{\rm var}$, $F_8$, and X$_0$ alone. 

On the other hand, we do find that targets on the flicker floor show a clear gradient in FUV verses R$_{\rm var}$, which ultimately we have used to illustrate a gradient in $\sigma_{RV}'$. As the flicker floor is presumed to trace log g, it is likely that this gradient in $\sigma_{RV}'$ is due to increased granulation noise as stars evolve along the main sequence. We also found a distinct difference in the level of measured photometric noise and predicted $\sigma_{RV}'$ for those stars hotter than 6000~K as compared to the cooler stars. We attribute the larger astrophysical noise in the hotter stars to the presence of larger, brighter, longer-lived granules with faster velocities and greater photometric contrast \citep{beeck13}. 

The approach outlined in this paper illustrates how the use of photometric measurements (R$_{\rm var}$, $F_8$ and X$_0$) can serve as a proxy to estimate RV noise for magnetically quiet stars, thus creating a filter mechanism for optimal RV follow-up, and subsequently trace their stellar evolution. Future work could greatly improve the accuracy and precision in the approach presented here to make such a filter mechanism a truly viable option for a wide variety of stellar hosts. The presence of binaries should be examined in future to determine the degree to which they contaminate the results. The effects of extinction should also be investigated as this may be an additional source of scatter in the relationships explored in this paper (the extinction values used in this work are from the KIC for the B-V colour). Knowledge of inclination angles in the future could also greatly help constrain the large parameter space for targets above the flicker floor. Moreover, the $\sigma_{RV}'$ estimations could be improved by obtaining bolometric fluxes and photospheric FUV predictions from atmospheric modelling to obtain normalised FUV excess values. This would in turn allow us to convert directly from FUV excess to ${R'}_{HK}$, which is preferred over converting from FUV emission alone since UV excess is the stronger stellar activity indicator. The FUV excess derivation from \cite{smith10} could also be explored in future work, where the authors explicitly incorporate ${R'}_{HK}$ into their definition of FUV excess. Additional improvements could be made by obtaining high precision observational measurements for both ${R'}_{HK}$ and of course RV. 

Nonetheless, for quiet stars in our sample we have established fairly well-defined relationships between R$_{\rm var}$, $F_8$, X$_0$ and estimated $\sigma_{RV}'$, and found these to be in agreement with solar observations. Such relationships are very important as these quiet stars are the best candidates for planet confirmation in general, but in particular they are the best hope for the confirmation of habitable Earth-like worlds. Furthermore, as photometrically quiet stars can still exhibit astrophysical noise up to $\sim$ 10 m s$^{-1}$, it is prudent to know which these stars are also spectroscopically quiet if we hope to confirm long-period, low-mass terrestrial planets in the future.

\section*{\sc Acknowledgments}
This research has made use of NASA's Astrophysics Data System Bibliographic Services. The authors would like to acknowledge the use of Filtergraph (http://filtergraph.vanderbilt.edu) for initial brainstorming and analysis. HMC acknowledges support from a Queen's University Belfast university scholarship and funding from the Sigma Xi Grants-in-Aid of Research. CAW would like to acknowledge support by STFC grant ST/I001123/1 and the Leverhulme Trust. FAB acknowledges support from a NASA Harriet Jenkins Fellowship and a Vanderbilt Provost Graduate Fellowship. We thank the anonymous referee both for their comments which have helped improve the content and clarity of this paper, as well as for their enthusiasm for the work presented here.

\bibliographystyle{apj}
\bibliography{abbrev,mybib}

\end{document}